\documentclass{article}

\usepackage{color}
\usepackage{latexsym}
\usepackage{amssymb}
\usepackage{gensymb}
\usepackage{subfigure} 
\usepackage{booktabs}
\usepackage{diagbox}
\usepackage{multirow}
\usepackage{array}
\usepackage{threeparttable}

\usepackage{cite}
\usepackage{graphicx}

\usepackage{geometry}
\usepackage{amsmath}

\usepackage{algorithmic}

\usepackage{array}

\usepackage{authblk}
\usepackage{textcomp}
\usepackage{color}
\usepackage[misc]{ifsym} 

\begin{document}

\title{Field-free Deterministic Magnetization Switching Induced by Interlaced Spin-Orbit Torques}
\author[1,2]{Min Wang}
\author[1,2,3,*]{Zhaohao Wang }
\author[1,4]{Chao Wang}
\author[1,2,3,*]{Weisheng Zhao }
\affil[1]{Fert Beijing Research Institute, Beihang University, Beijing, China}
\affil[2]{School of Microelectronics, Beihang University, Beijing, China}
\affil[3]{Beijing Advanced Innovation Center for Big Data and Brain Computing, Beihang University, Beijing, China}
\affil[4]{School of Electronics and Information Engineering, Beihang University, Beijing, China} 
\affil[*]{zhaohao.wang@buaa.edu.cn and weisheng.zhao@buaa.edu.cn}
\date{}
\maketitle

\begin{abstract}
Spin-orbit torque (SOT) based magnetic random access memory (MRAM) is envisioned as an emerging non-volatile memory due to its ultra-high speed and low power consumption. The field-free switching schema in SOT devices is of great interest to both academia and industry. Here we propose a novel field-free deterministic magnetization switching in a regular magnetic tunnel junction (MTJ) by using two currents sequentially passing interlaced paths, with less requirements of manufacturing process or additional physical effects. The switching is bipolar since the final magnetization state depends on the combination of current paths. The functionality and robustness of the proposed schema is validated through both macrospin and micromagnetic simulation. The influences of field-like torque and  Dzyaloshinskii-Moriya interaction (DMI) effect are further researched. Our proposed schema shows good scalability and is expected to realize novel digital logic and even computing-in-memory platform.

\end{abstract}

\section{Introduction}

Fast and energy-efficient switching of the perpendicular magnetization has attracted intensive research interests since it directly influences the application potential of spintronics devices such as magnetic tunnel junction (MTJ)\cite{apalkov2016magnetoresistive,bhatti2017spintronics,cao2018memory}. Currently, the spin transfer torque (STT), as the mainstream technology in non-volatile memory, is suffering from the bottlenecks of speed, energy and reliability\cite{slonczewski1996current,berger1996emission,brataas2012current,khvalkovskiy2013basic}. For instance, the intrinsic incubation delay of the STT limits the speed of magnetization switching. The charge current for inducing the STT directly flows through the MTJ and hence easily damages the tunnel barrier. To overcome these bottlenecks, the emerging spin orbit torque (SOT) technology has been proposed and shows the advantages of faster speed, lower energy, and higher reliability\cite{ramaswamy2018recent,liu2012spin,cubukcu2014spin,miron2011fast,fukami2016spin,lee2016emerging,wang2018progresses,huang2020recent,zhao2018asymmetric}. Nonetheless, for achieving the deterministic switching of perpendicular magnetization, most of the SOT prototype devices have to be used in conjunction with an additional magnetic field, which blocks the practical realization of the SOT-based circuits and architectures. Therefore, field-free SOT switching schema is drawing more and more attention on account of practical applications. 

Up to now, various methods of field-free SOT-induced switching have been reported, such as creating a lateral structural asymmetry via special fabrication process\cite{yu2014switching,akyol2015current}, tilting the anisotropy \cite{you2015switching}, adopting the antiferromagnetic layer with strong spin orbit coupling \cite{fukami2016magnetization,Lau2016Spin,oh2016field,van2016field}, adjusting Dzyaloshinskii-Moriya Interaction (DMI) effect \cite{chen2019field,dai2020field,wu2020deterministic} or field-like torque \cite{legrand2015coherent}, introducing shape anisotropy \cite{Wang2019Field}, etc \cite{sverdlov2019two,go2018switching,de2020two,safeer2016spin,deng2018electric,kazemi2016all,wang2018field,sato2018two,cai2017electric}. However, these methods bring out special requirements in fabrication process or material type of the device. In addition, for \cite{chen2019field,legrand2015coherent}, the magnetization switching is unipolar, which means that a read-before-write operation is required in the circuit-level application.  

In this paper, we propose a novel SOT-based technology for achieving field-free deterministic switching of perpendicular magnetization. Our proposal is validated by both macrospin and micromagnetic simulation. Compared with the existing methods, our proposal could be implemented in a regular SOT-MTJ, without the need of special fabrication process or physical effects. Furthermore, our proposal leads to bipolar switching, which could be naturally applied in the spintronics memories or computing circuits. 

\section{Model and simulation}

\begin{figure}[h]
	\centering
	\includegraphics[width=10cm]{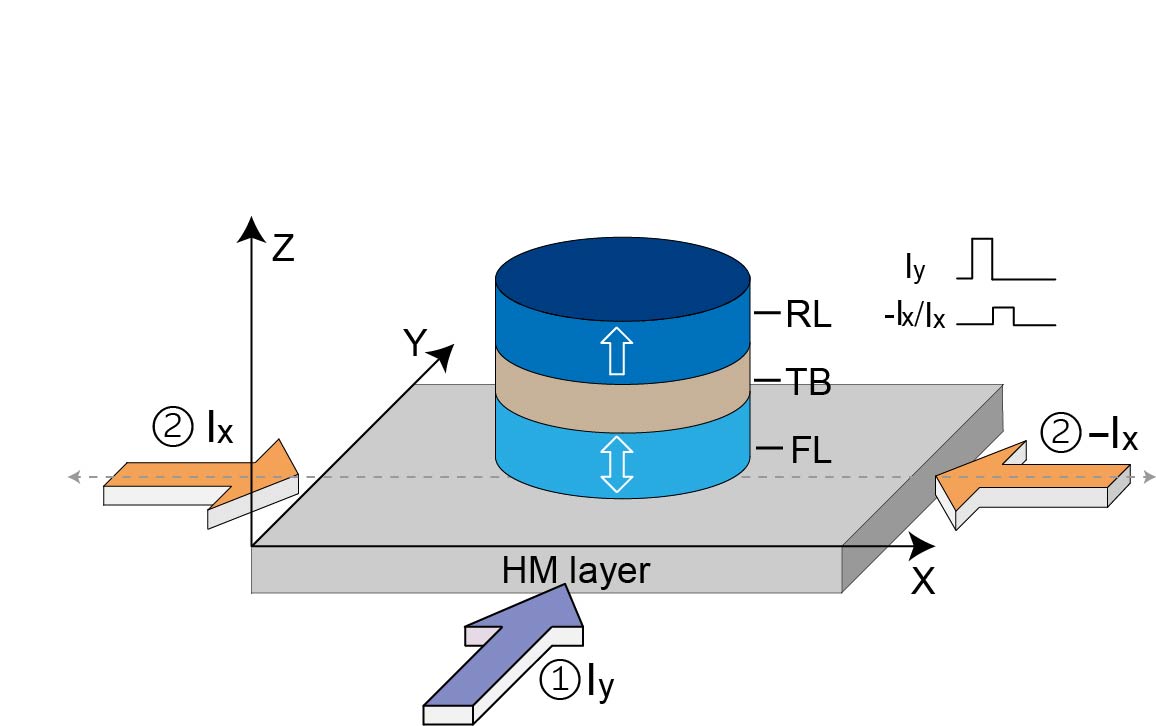}
	\quad
	\caption{Schematic structure of the studied device, including free layer (FL)/tunnel barrier (TB)/reference layer (RL), and heavy metal (HM) layer. The current could be applied along $x$ or $y$ direction.}
	\label{1}
\end{figure}

The schematic structure of our proposed field-free device is illustrated in Fig. \ref{1}. A regular MTJ is utilized, including a tunnel barrier (TB) sandwiched between two ferromagnetic layers, i.e. free layer (FL) and reference layer (RL), fabricated above a heavy metal layer. The magnetization of RL is fixed while the magnetization of FL can be switched by current-induced SOT through heavy metal (HM) layer. Our schema requires two currents paths, i.e, $\pm x$ axis and $\pm y$ axis, which can be achieved by adding four electrodes to the HM layer. The magnetization dynamics of the FL can be charactered by a modified Landau-Lifshitz-Gilbert (LLG) equation.

\begin{equation}
\frac{\partial {\mathbf{m}}}{\partial t} = -\gamma \mu_0 \mathbf{m}\times \mathbf{H_{eff}}  +\alpha\mathbf{m}\times\frac{\partial \mathbf{m}}{\partial t}+\boldsymbol{\tau_{DL}+\boldsymbol{\tau_{FL}}}
\end{equation}
\begin{equation}
\boldsymbol{\tau_{DL}}=J_{sot}\xi\lambda_{DL}\mathbf{m}\times({\boldsymbol{\sigma}}\times\mathbf{m}) 
\end{equation}
\begin{equation}
\boldsymbol{\tau_{FL}}=-J_{sot}\xi\lambda_{FL}(\mathbf{m}\times{\boldsymbol{\sigma}})
\end{equation}
\begin{equation}
\xi=\frac{\gamma\hbar}{2et_FM_s}
\end{equation} 
where $\gamma$ is the gyromagnetic ratio. $\mu_0$ refers to the vacuum permeability. $\sigma$ denotes the polarisation vector of the SOT-induced spin injection. $J_{sot}$ is the SOT current density. $\xi$ is a device-dependent parameter.  $\lambda_{DL}$ and $\lambda_{FL}$ reflect the magnitudes of damping-like torque (DLT) and field-like torque, respectively. $\lambda_{DL}$, is equivalent to the spin Hall angle. Other parameters are listed in Table \ref{para}, which is consistent with the state-of-the-art technologies.

\begin{table*}[h]
	\renewcommand{\arraystretch}{1.3}
	\caption{Parameters of simulation }
	\label{para}
	\centering
	\begin{tabular*}{1\textwidth}{@{\extracolsep{\fill}}llll}
		\hline
		Symbol & Paramter & Default value &Ref.\\
		\hline
		-       & MTJ area                 &$0.25\times\pi\times70~{\rm nm}\times70~{\rm nm}$\\
		$t_F$   &Thickness of FL           &$0.8~{\rm nm}$\\
		$\alpha$&Damping constant          &$0.3$ &\cite{metaxas2007creep} \\
		$A$     & Exchange constant        & $1\times10^{-11}~{\rm J/m}$ \\
		$K_u$   & Magnetic anisotropy      & $8\times10^5~{\rm J/m^3}$ &\cite{chen2019field} \\
		$M_s$   & Saturation magnetization &$1.1\times10^6~{\rm A/m}$ &\cite{chen2019field}\\
		$\theta_{sot}$&Spin Hall angle &$0.3$ &\cite{liu2015correlation,pai2012spin}\\	
		\hline
	\end{tabular*}
\end{table*}

 The combinations of current \{$+I_y$, $+I_x$\} and \{$+I_y$, $-I_x$\} are chosen to achieve bipolar switching, as shown in Fig. \ref{1}. An exemplary switching process can be accomplished in two steps. 
 
 $\mathbf{Step 1.}$ A charge current in $+y$ direction ($+I_y$), with current density $J_{sot1}$ and pulse width $t_{p1}$, is applied to the heavy metal layer and induces a spin current polarized along $-x$ direction. The normalized magnetization vector ($\mathbf{m}$) is pulled to the $-x$ direction and hence the metastable in-plane state, regardless of initial magnetization. 
 
\begin{figure*}[h]
	\centering
	\includegraphics[width=18cm]{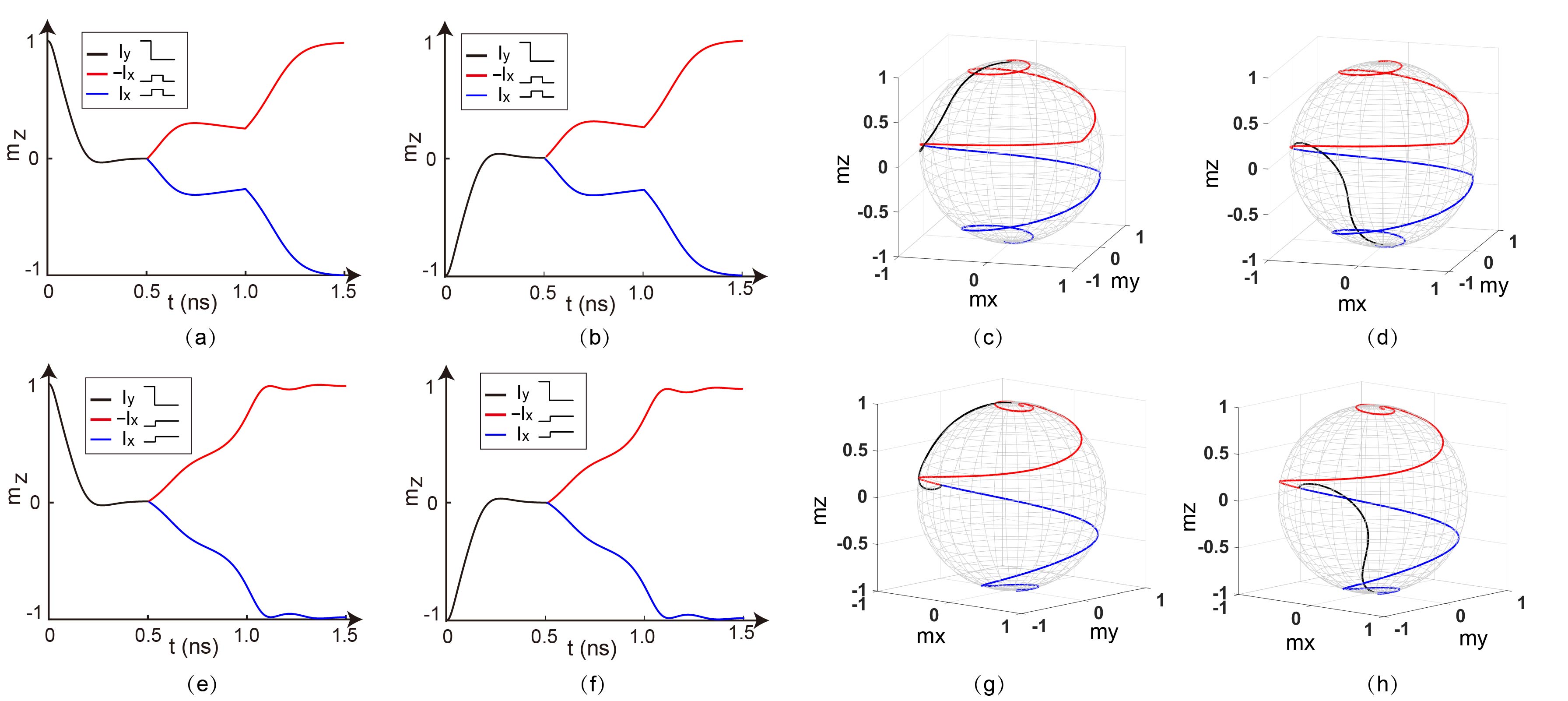}
	\quad
	\caption{(a)(b)(e)(f) Typical simulation results of time-dependent z-component magnetization ($m_z$) for (a)(b) Mode-I switching and (e)(f) Mode-II switching. (c)(d)(g)(h) Trajectories of the magnetization vector for (c)(d) Mode-I switching and (g)(h) Mode-II switching.}
	\label{SW1}
\end{figure*}

 $\mathbf{Step 2.}$ $+I_y$ is removed and subsequently the another current ($+I_x$ or $-I_x$) is applied in $+x$ or $-x$ direction, with current density $J_{sot2}$ and pulse width $t_{p2}$. Starting from metastable in-plane state, the magnetization switching could be achieved. Depending on the magnitude of $J_{sot2}$, two modes of magnetization switching could occur. First, if $J_{sot2}$ is large enough, the magnetization firstly rotates towards $-z$ or $+z$ axis under the action of $+I_x$ or $-I_x$, and then returns to in-plane direction. Therefore, the deterministic switching could be achieved as long as $+I_x$ or $-I_x$ is turned off before the magnetization significantly returns to in-plane direction. This mechanism is named Mode-I switching. The typical curves of time-dependent $m_z$ and trajectories of the magnetization vector are shown in Fig. \ref{SW1}(a)-(d). Second, while $J_{sot2}$ is relative weak, the magnetization could directly be switched to $-z$ or $+z$ direction, even if the $+I_x$ or $-I_x$ is not turned off. This mechanism is named Mode-II switching, as shown in Fig. \ref{SW1}(e)-(h). It is important to mention that both Mode-I and Mode-II switching allow the timing overlap of two current pulses, showing good feasibility in circuit-level application. Hereinafter, the current pulses applied in steps 1 and 2 are referred to as "the first current" and "the second current", respectively.
 
 Below, firstly we will focus on the influences of the current pulse parameters on the magnetization switching. Afterwards, the robustness of our proposed schema is studied. Finally, micromagnetic simulation is performed to reveal more features of the magnetization switching process.

\section{Results and discussion}
The key point of our proposal is to utilize the magnetization precession caused by the second current. It is worthy mentioning that the choice of current paths is not unique since the studied device has structure symmetry between $x$ and $y$ axes. Table \ref{final state} lists the feasible combinations of the current pulses for the bipolar switching. It is observed that when the second current is in a clockwise (counterclockwise) relationship with the first current, magnetization will be switched to $-z$ ($+z$) direction. In practical applications, the current combination can be flexibly chosen, which provides the potential as logic devices and computing circuits.

\begin{table*}[t]
	\small
	\caption{The feasible combination of current pulses}
	\label{final state}
	\begin{tabular*}{1\textwidth}{@{\extracolsep{\fill}}lll}
		\hline
		The first current & The second current & Final state of $m_z$\\
		\hline
		$+I_y$  & $+I_x$  &$-1$\\
		$+I_y$  & $-I_x$  &$+1$\\
		$-I_y$  & $+I_x$  &$+1$\\
		$-I_y$  & $-I_x$  &$-1$\\
		$+I_x$  & $+I_y$  &$+1$\\
		$+I_x$  & $-I_y$  &$-1$\\
		$-I_x$  & $+I_y$  &$-1$\\
		$-I_x$  & $-I_y$  &$+1$\\
		\hline
	\end{tabular*}
\end{table*}

Considering the symmetry among the various combinations, we will only analyse the case of \{step 1: $+I_y$; step 2: $-I_x$\} in this section and take the $\emph{-z-to-+z}$ switching process as example. In addition, as the first current is only responsible for driving the magnetization to in-plane direction, we will mainly discuss the influence of the second current. Macrospin simulation is performed for the preliminary analysis. Then the domain wall dynamics and DMI effect are discussed through micromagnetic simulation.

\subsection{Influence of the second current}

\subsubsection{Optimal current pulse width for Mode-I switching}

\begin{figure}[h]
	\centering
	\subfigure[]{\includegraphics[width=5cm]{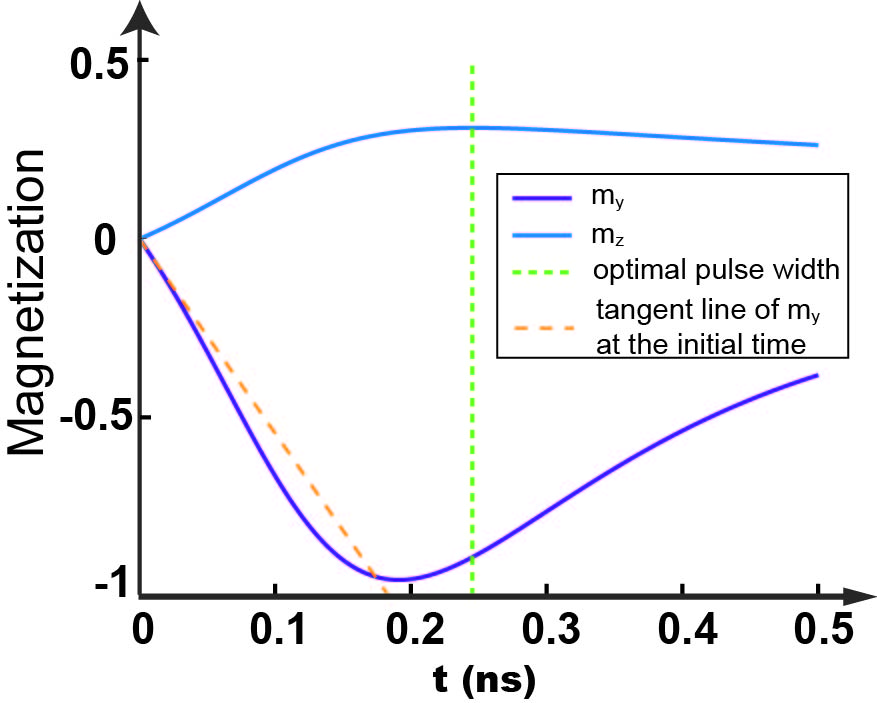}}
	\quad
	\subfigure[]{\includegraphics[width=2cm]{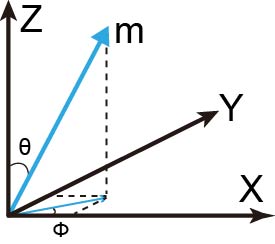}}
	\quad
	\subfigure[]{\includegraphics[width=4cm]{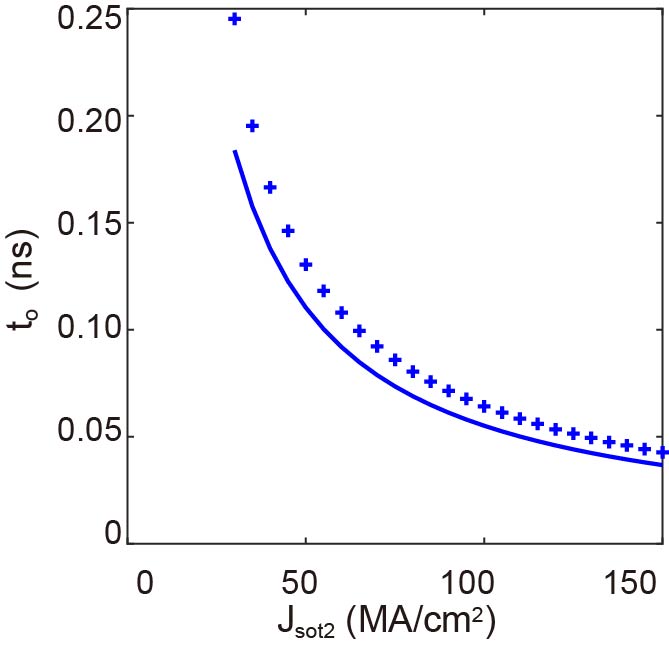}}
	\quad
	\subfigure[]{\includegraphics[width=4cm]{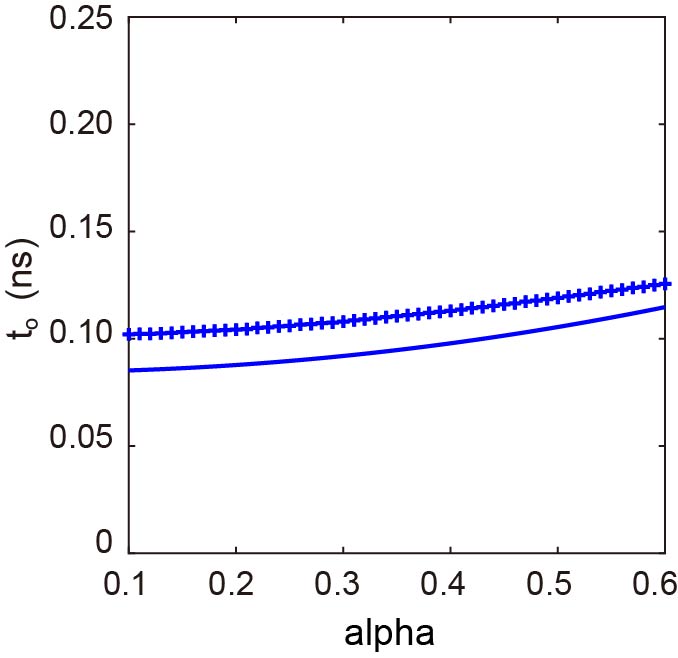}}
	\caption{(a) Time evolution of $m_y$ and $m_z$ under the action of $-I_x$. (b) Coordinate system for our study. (c) The theoretically-estimated $t_o$ (solid line) and simulated $t_o$ (points) as a function of current density. (d) The theoretically-estimated $t_o$ (solid line) and simulated $t_o$ (points) as a function of $\alpha$.}
	\label{3}
\end{figure}

The second current ($-I_x$ in our example) plays a key role in the magnetization switching. As shown in Fig. \ref{SW1}(a)-(d), for Mode-I switching, $-I_x$ must be removed at an appropriate time, otherwise the magnetization vector will be pulled to in-plane direction again by the SOT. In order to obtain the optimal pulse width, we show the curves of $m_y$ and $m_z$ during step 2 in Fig. \ref{3}(a). It is observed that the intersection between the tangent line of $m_y$ at the initial time and the time axis is close to the optimal pulse width (i.e. corresponding to the extreme value of $m_z$ ($m_{z,ex}$), see Fig. \ref{3}(a)). Through this method, the estimated value of the optimal pulse width ($t_o$) could be derived by solving the LLG equation in spherical coordinate (see Fig. \ref{3}(b)), as follows

\begin{equation}\label{dmy0}
\frac{\partial m_y}{\partial t}\bigg|_{\theta_0=\pi/2, \phi_0=\pi} = -\frac{J_{sot}\xi\theta_{sot}}{1+\alpha^2}
\end{equation}
\begin{equation}\label{t_o}
t_o=\frac{1+\alpha^2}{J_{sot}\xi\theta_{sot}}
\end{equation}

Figure \ref{3}(c) presents the results of the theoretically-estimated $t_o$ (solid line) and simulated $t_o$ (points), where the accuracy is improved as current density increases. Furthermore, Eq. (\ref{t_o}) correctly describes the parameter dependence of $t_o$, as shown in Fig. \ref{3}(c) and (d). A larger $\alpha$ can enlarge $t_o$ and allows better tolerance to the time deviation.

\subsubsection{Damping constant}
\begin{figure}[htbp]
	\centering
	\subfigure[]{\includegraphics[width=6cm]{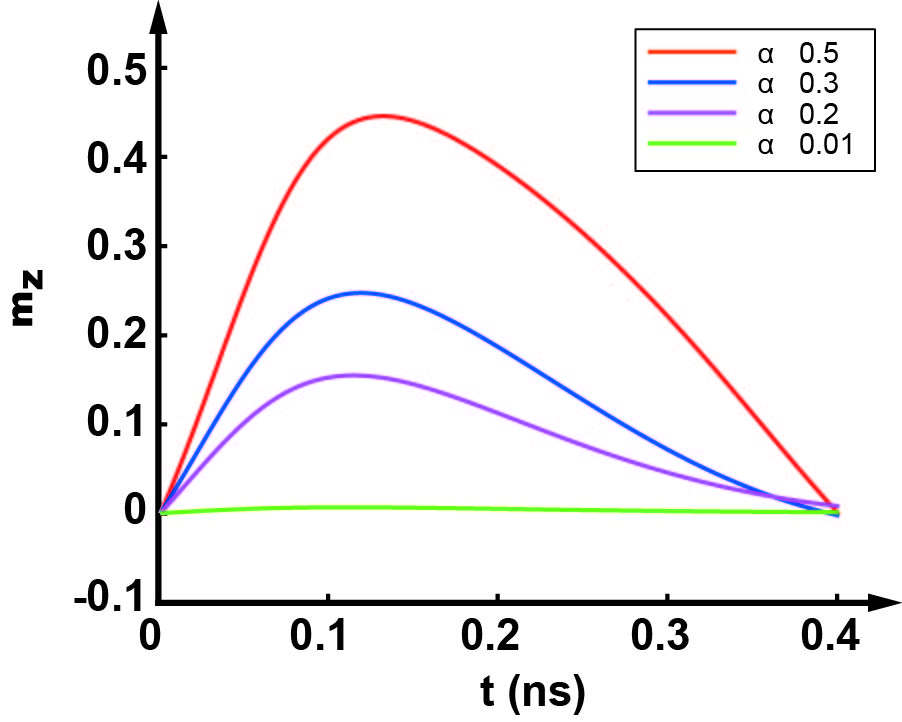}}
	\quad
	\subfigure[]{\includegraphics[width=6cm]{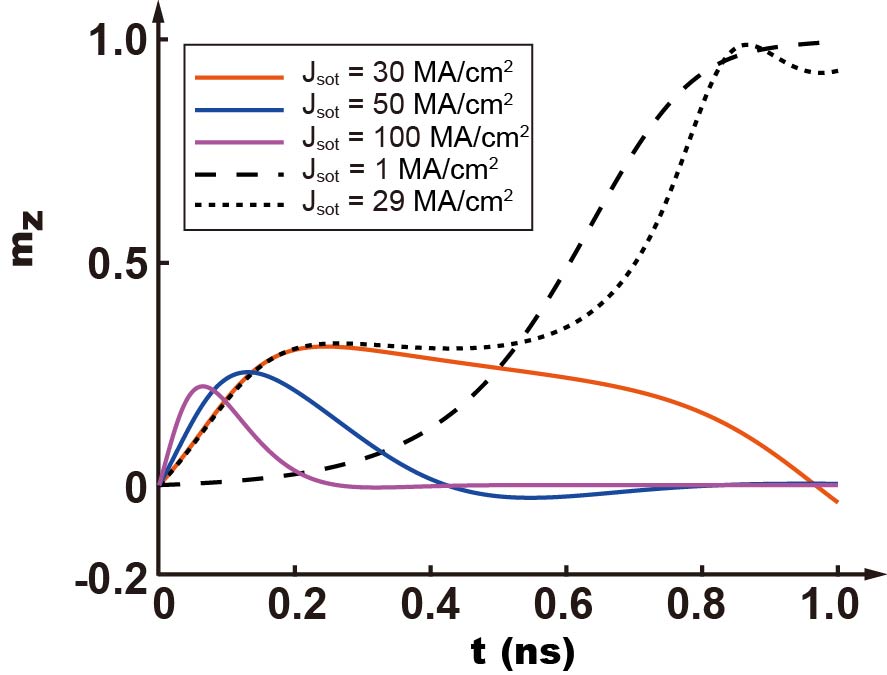}}
	\quad
	\caption{Influence of (a) $\alpha$ and (b) $J_{sot}$ on the magnetization switching. (a) $J_{sot}=55~{\rm MA/cm^2}$. (b) $\alpha=0.3$.}
	\label{4}
\end{figure}

In addition to the positive effect on $t_o$, larger $\alpha$ can improve the reliability of Mode-I switching, as shown in Fig. \ref{4}(a). The reliability can be reflected by the initial slope of $m_z$, estimated by

\begin{equation}\label{dmz0}
\begin{aligned}
\frac{\partial m_z}{\partial t} \bigg|_{t=0} =  \frac{\alpha J_{sot}\xi\theta_{sot}}{1+\alpha^2}
\end{aligned}
\end{equation}

\subsubsection{Critical current density}
As mentioned above, the second current density ($J_{sot2}$) has a dramatic effect on the process of magnetization switching. For the Mode-II switching, the current is so weak that it cannot pull the magnetization back into the plane again, the magnetization will directly approach the +z axis, as the dotted lines in Fig. \ref{4}(b). For the Mode-I switching, the curve of $m_z$ shows a extreme value, as solid lines in Fig. \ref{4}(b). In this case, the $J_{sot2}$ dependence of the $m_z$ curve could be described by the above Eqs. (\ref{t_o}) and (\ref{dmz0}). The critical current density for the boundary between Mode-I and Mode-II switching is defined as $J_{o}$. Supposed the boundary state, where $m_{z, ex}\approx \alpha$ and the $\partial m_z/\partial t=0$, thus, the estimated $J_{o}$ calculated as,

\begin{equation}\label{Jo}
J_o = \frac{\alpha\gamma\mu_0\sqrt{1-\alpha^2} H_{eff}}{\sqrt2\xi\theta_{sot}}.d
\end{equation}
where $d=1.1$ is a fitting factor. 

According to the mathematical meaning, the product of Eqs. (\ref{t_o}) and (\ref{dmz0}) is $m_{z, ex}$, numerically equals to $\alpha$. This conclusion is in good agreement with Fig. \ref{4}(a). Nevertheless, there still exists deviation between $m_{z, ex}$ and $\alpha$, as can be seen from Fig. 4(a), which is attributed to the fact that $m_{z, ex}$ is also related to the current density, as shown in Fig. \ref{4}(b). This deviation could be minimized while the applied current density is set to $J_o$.

\subsubsection{Angle between two current paths}

\begin{figure}[htbp]
	\centering
	\includegraphics[width=15cm]{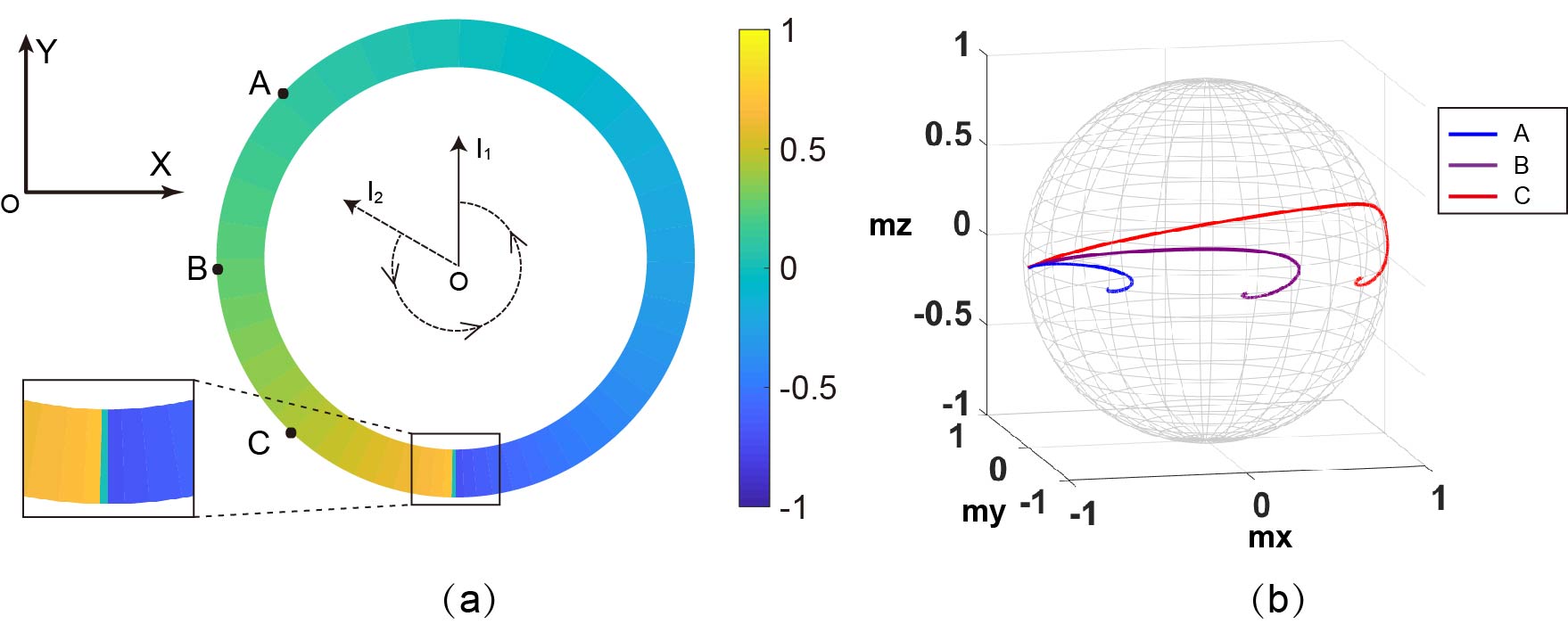}
	\quad
	\caption{(a) The results of $m_{z, ex}$ by sweeping the angle between two current paths. The first current is in $+y$ direction, while the direction of second current changes according to the coordinate axis.  (b) Typical trajectories of the magnetization vector. Points A, B and C correspond the second current at an anticlockwise angle of 45$\degree$, 90$\degree$ and 135$\degree$ with the first current, respectively.}
	\label{CT}
\end{figure}

In addition to the damping constant and current density, the angle between two current paths could be changed to improve the reliability of the magnetization switching. Figure \ref{CT}(a) presents $m_{z,ex}$ as a function of the angle between two current paths, where the colorbar presents $m_{z,ex}$. Here we fix the direction of the first current ($I_1$), while change the direction of the second current ($I_2$). It is seen that the absolute value of $m_{z,ex}$ monotonically increases while the angle between two current paths increases. More comprehensive trajectories of the magnetization vector could be seen in Fig. \ref{CT}(b). Therefore, two current paths could be designed in an interlaced way, not limited to the vertical relationship. Note that an abrupt change of $m_{z, ex}$ occurs in the opposite direction of $I_1$ (i.e. $-y$ direction), which is the boundary of clockwise and counterclockwise angle between two current paths, as well as the boundary of rotation direction.

\newpage
\subsection{Thermal robuteness}
Following, the thermal fluctuation is considered to evaluate the robustness of the proposed schema. We focus on the influences of pulse width, current density, and field-like torque.

\subsubsection{Current density and pulse width}
\begin{figure*}[htbp]
	\centering
    \quad
    \includegraphics[width=18cm]{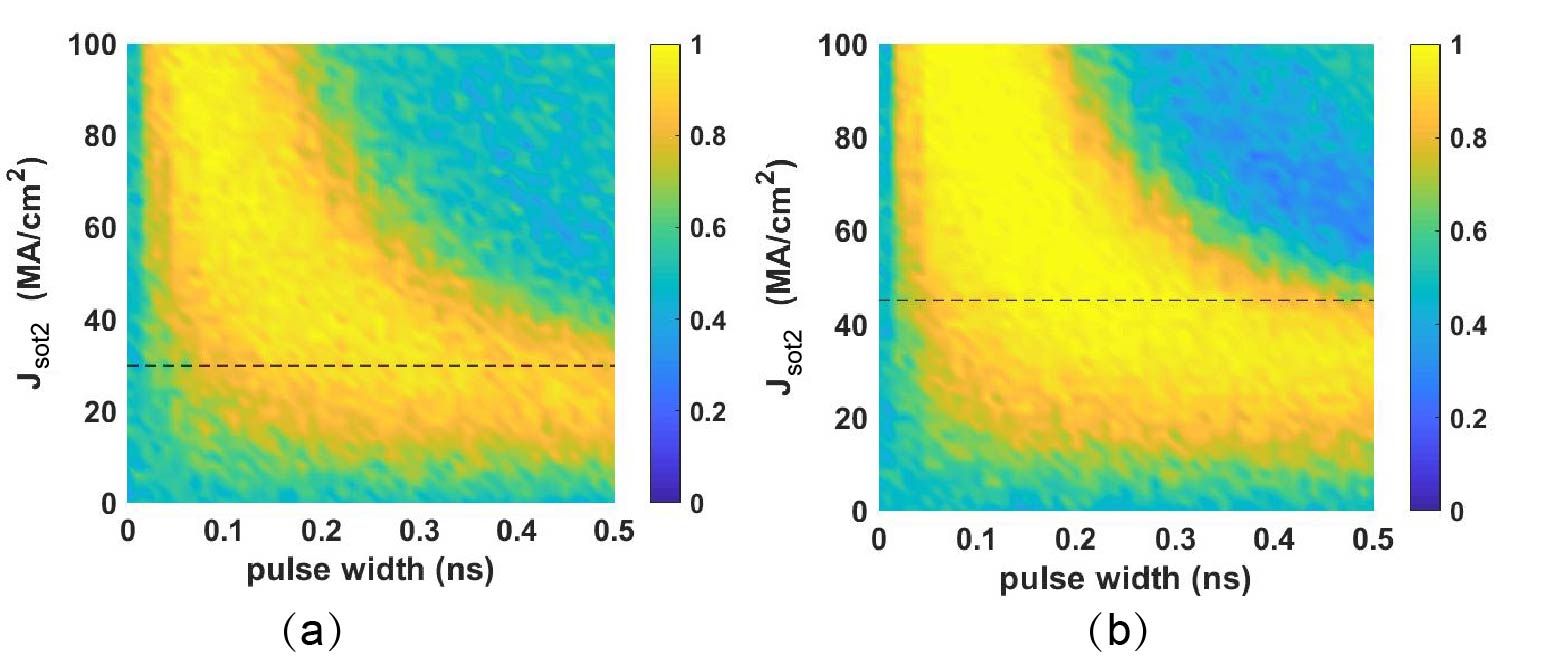}
	\caption{Switching probability in down-to-up process as a function of the second current density and pulse width. Each pixel means 100 trials. (a) $\alpha = 0.3$, (b) $\alpha = 0.5$. Dots lines correspond to critical current density values in Eq. (\ref{Jo}), with $d=1.1$.}
	\label{Psw}
\end{figure*}

Simulation results in Fig. \ref{Psw} (a) illustrates the switching probability as a function of current density and pulse width, at various damping constants. As expected, a current window for the robust switching (probability is close to 1) could be seen from both figures. This window could be well explained by the above analysis. The smaller current induces an insufficient torque which fails to drive the magnetization switching. The excessively large current tends to pull the magnetization to metastable in-plane direction and deteriorates the robustness. The current window for the robust switching could be enlarged by increasing the damping constant, since larger damping constant leads to larger $m_{z, ex}$.

\subsubsection{Influence of field-like torque}
\begin{figure*}[htbp]
	\centering
	\includegraphics[width=15cm]{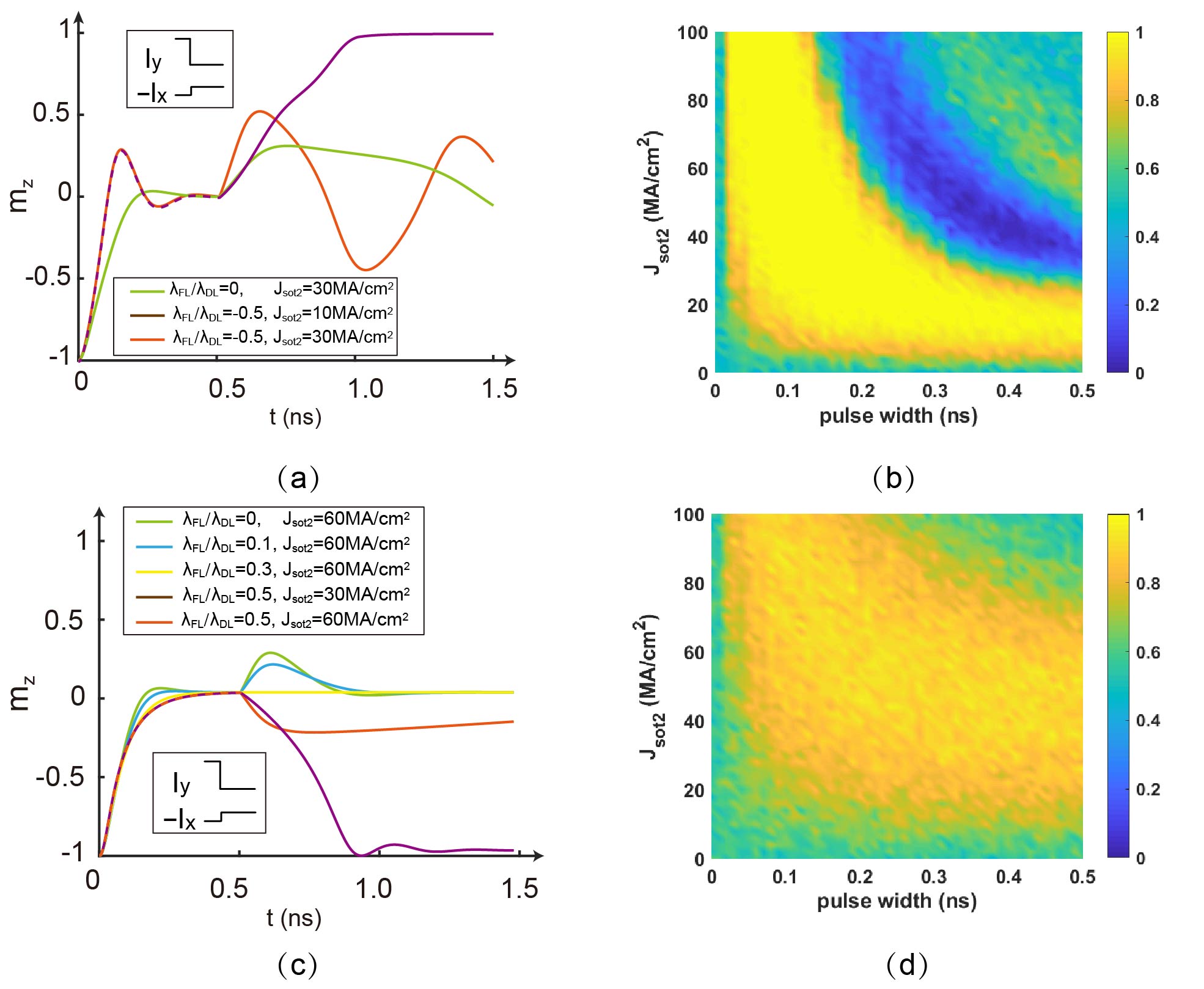}
	\quad
	\caption{The influences of (a)(b) negative field-like torque and (c)(d) positive field-like torque. (a)(c) time-dependent $m_z$. (b)(d) Switching probability as a function of the second current density and pulse width.}
	\label{FLT-MATLAB}
\end{figure*}

In this subsection, the influence of field-like torque on the magnetization switching is discussed. We found that Mode-II switching could still occur in the presence of non-zero field-like torque, as the violet lines shown in Fig. \ref{FLT-MATLAB}(a) and (c). The profile of Mode-II switching curve is not significantly influenced by the field-like torque. Below we focus on the case of Mode-I switching.

At zero temperature, a negative field-like torque could cause the precession of magnetization around the in-plane direction (see the red line of Fig. \ref{FLT-MATLAB}(a)). Thus, the deterministic magnetization switching could be completed if the pulse width is set to an appropriate value. This inference is validated by Fig. \ref{FLT-MATLAB}(b) which shows the results of switching probability at room temperature. A window of pulse width for the robust switching is clearly seen. In addition, another window of pulse width for zero-probability switching is also seen, which is consistent with the precession behaviours shown in Fig. \ref{FLT-MATLAB}(a).

A positive field-like torque could decrease the $m_{z, ex}$ and even reverse the polarity of magnetization switching, as shown in Fig. \ref{FLT-MATLAB}(c). As to the switching probability at room temperature, the current window for robust switching is expanded, as indicated by the comparison between Fig. \ref{FLT-MATLAB}(d) and Fig. \ref{Psw}(a). However, the switching probability in Fig. \ref{FLT-MATLAB}(d) is smaller than that in Fig. \ref{Psw}(a), due to the decrease of $m_{z, ex}$.

\newpage
\subsection{Mircomagnetic simulation}
\begin{figure*}[htbp]
	\centering
    \includegraphics[width=15cm]{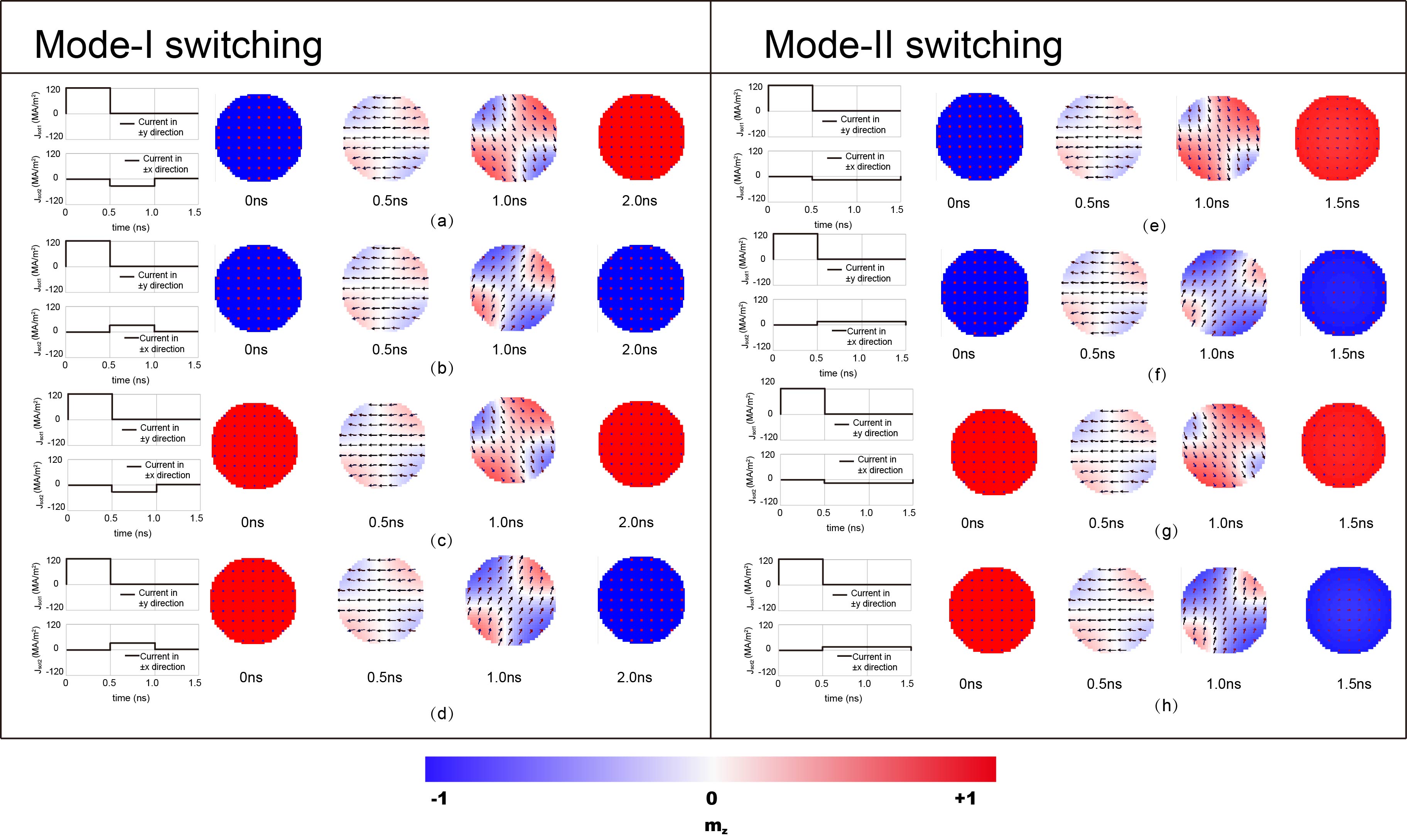}
	\quad
	\caption{Snapshots of time-dependent micromagnetic configurations for Mode-I and Mode-II switching.}
	\label{DMI1}
\end{figure*}

Micromagnetic simulation is performed with OOMMF\cite{donahue1999oommf} to reveal more details about the magnetization switching. The mesh size is set to $2~nm\times2~nm\times0.8~nm$. Typical results are shown in Fig. \ref{DMI1}, where it can be seen that the magnetization switching is completed through the domain wall dynamics. At the end of the first current, the magnetic domains are configured to be a balance state ($<$$m_z$$>$=0). The second current is responsible for the deterministic switching. Both Mode-I and Mode-II switching are achieved, in agreement with Table II.

\subsubsection{Critical current in micro-simulation}

\begin{figure}[htbp]
	\centering
	\includegraphics[width=8cm]{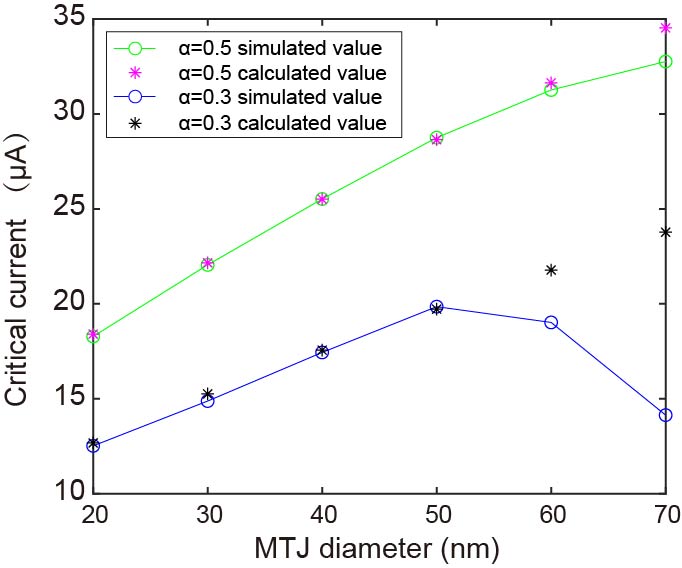}
	\quad
	\caption{The critical current as a function of the MTJ diameter. Here the calculated value is obtained based on Eq. (\ref{Jo}).}
	\label{IJ}
\end{figure}

Figure \ref{IJ} depicts the micromagnetic simulation results of the critical current versus the MTJ diameter, showing potential of continuous scaling. By properly setting the fitting factor $d$ in Eq. (\ref{Jo}), the theoretical value of critical current is generally in agreement with the micromagnetic simulation results. However, for $\alpha = 0.3$, the deviation occurs when the MTJ diameter is larger, which may be attributed to the transformation from uniform switching to domain-governed switching.

\subsubsection{DMI effect}
\begin{figure*}[htbp]
	\centering
    \includegraphics[width=16cm]{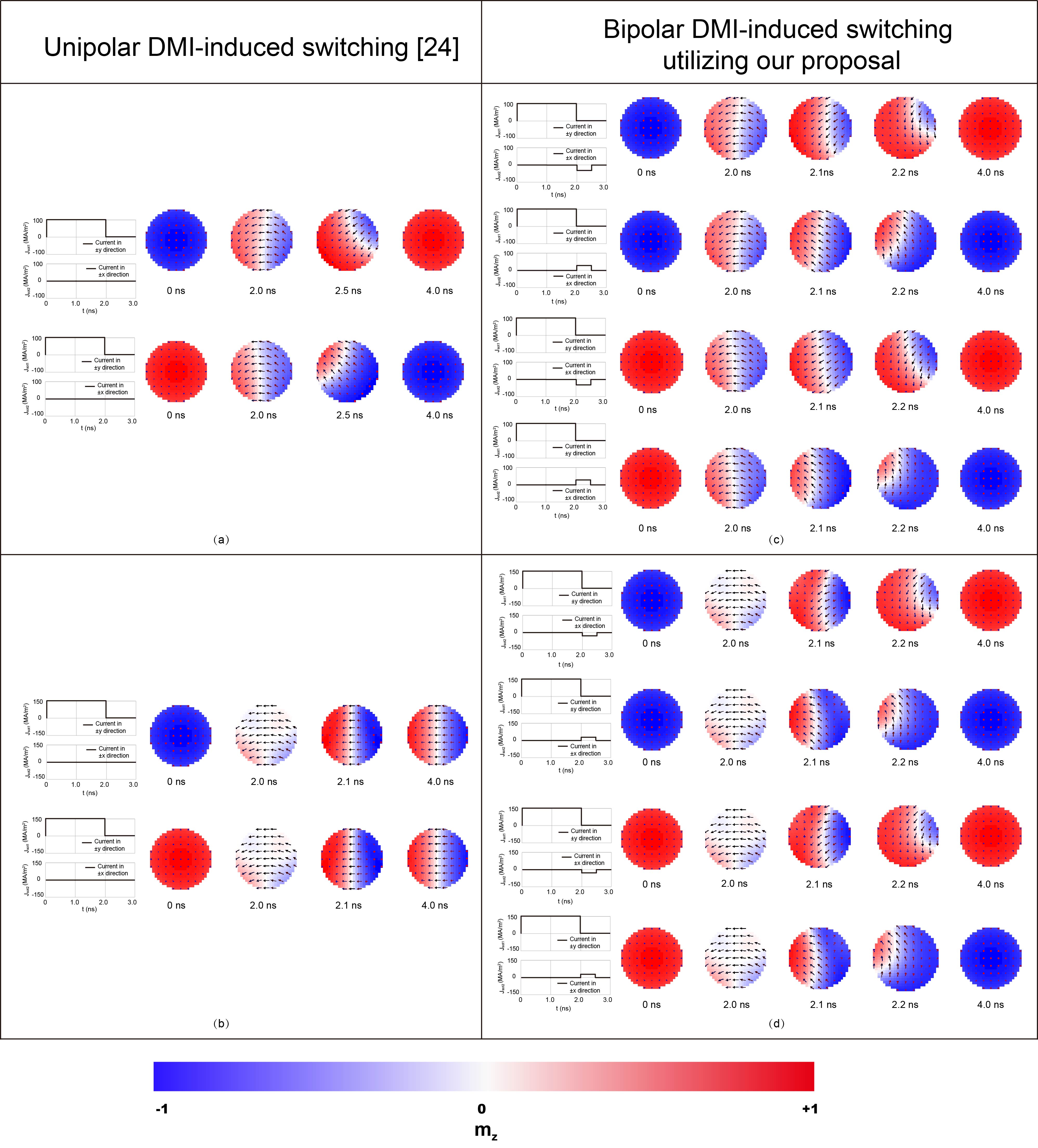}
	\quad
	\caption{Snapshots of magnetization during unipolar DMI-induced switching \cite{chen2019field} and bipolar DMI-induce switching utilizing our proposal. }
	\label{DMI2}
\end{figure*}

Recently, it was demonstrated that the combination of DMI effect and SOT can lead to field-free magnetization switching \cite{chen2019field}. The corresponding simulation results are shown in Fig. \ref{DMI2}(a)(b), where the current-induced SOT drives the domain wall across the equilibrium position and hence achieves deterministic switching in the presence of DMI effect. However, the switching is unipolar, which is not beneficial to the real circuit-level application since a read-before-write operation is required. In addition, the available window of current density for the deterministic switching is limited.

In contrast, our schema leads to bipolar deterministic switching even in the presence of DMI effect, as shown in Fig. \ref{DMI2}(c)(d). The reason is that the direction of domain wall motion depends on the polarity of the second current. Furthermore, our schema relaxes the limit of the window of the first current, since the final state is determined by the second current rather than the first current. 

\section{Conclusion}
We have proposed a novel schema of field-free deterministic magnetization switching by utilizing interlaced SOT currents. Depending on the magnitude of current density, two modes of bipolar switching were identified. The influences of key parameters on the magnetization switching were discussed through both macrospin and micromagnetic simulation. Theoretical models were deduced and shown good agreement with the simulation results. Compared with the existing field-free SOT schemata, our proposal could be implemented with a regular SOT-MTJ device, without the need of special process or physical effects, showing great application potential in non-volatile memory and logic computing.
\newpage
\bibliographystyle{IEEEtran}
\bibliography{reference}
\end{document}